\documentclass[12pt]{iopart}

\usepackage{amsbsy}
\usepackage{amssymb}
\usepackage{graphicx}
\usepackage{xcolor}
\usepackage{hyperref}
\begin{document}

\title[Work fluctuations for diffusion dynamics submitted to stochastic return]{Work fluctuations for diffusion dynamics submitted to stochastic return}

\author{Deepak Gupta$^1$ and Carlos A. Plata$^2$}

\address{$^1$ Department of Physics, Simon Fraser University, Burnaby, British Columbia V5A 1S6, Canada}
\address{$^2$ Física Teórica, Universidad de Sevilla, Apartado de Correos 1065, E-41080 Sevilla, Spain}
\ead{phydeepak.gupta@gmail.com}
\vspace{10pt}

\begin{abstract}
Returning a system to a desired state under a force field involves a thermodynamic cost, i.e., {\it work}. This cost fluctuates for a small-scale system from one experimental realization to another. We introduce a general framework to determine the work distribution for returning a system facilitated by a confining potential with its minimum at the restart location.
The general strategy, based on average over \textit{resetting pathways}, constitutes a robust method to gain access to the statistical information of observables from resetting systems. 
We exploit paradigmatic setups, where explicit computations are attainable, to illustrate the theory. Numerical simulations validate our theoretical predictions.
For some of these examples, a non-trivial behavior of the work fluctuations opens 
a door to
optimization problems. Specifically, work fluctuations can be minimized by an appropriate tuning of the return rate.
\end{abstract}

%
%
%
%
%

\section{Introduction}
\label{intro}
Stochastic resetting is a relatively recent, fast-growing
field within the realm of non-equilibrium statistical mechanics. 
Despite the simplicity of the concept, outstanding properties emerge when, on top of its natural dynamics, a stochastic system is submitted to a random reset to a certain state. Due to its rich phenomenology, stochastic resetting still represents an appealing test bench for researchers interested in non-equilibrium even a decade after its original modern formulation~\cite{11Evans}.  

Stochastic resetting exhibits remarkable theoretical properties that have made the former an excellent tool to analyze non-equilibrium stationary states~\cite{Pal_PRE, Sanjib_relax, 16Mendez, Gupta_2019}, thermodynamic speed limit~\cite{Deepak-SL}, thermodynamic uncertainty relation~\cite{21Pal}, inducing and optimizing the Markovian Mpemba effect~\cite{21Busiello}, antiviral therapies~\cite{Ramoso}, modelling cell division~\cite{Genthon_2022}, tax dynamics~\cite{Santra_2022}, just to cite a few 
hot topics in non-equilibrium statistical mechanics  
assisted by stochastic resetting. However, it is not only in the theoretical perspective that stochastic resetting becomes popular, there are certainly  a myriad of applications, from search and foraging processes~\cite{11Evans,11Evans_b,20Pal, Pal_FPT} to ecological disasters~\cite{20Plata} passing though enzymatic reactions~\cite{14Reuveni}, see~\cite{20Evans} for a thorough review.

In its original formulation~\cite{11Evans}, stochastic resetting was considered instantaneous, i.e., when the reset occurs the state of the system is instantaneously changed to a resetting state and the natural dynamics starts afresh. From  
a theoretical perspective, the \textit{simple} renovation of the dynamics allows to resort to a renewal formulation of the dynamics~\cite{17Roldan,18Chechkin}. Nevertheless, instantaneous reset is an ideal limit, which may be inconvenient to characterize real systems. For instance, a clear drawback of instantaneous resets is the thermodynamic cost. The cost required to drastically change the state of a physical state in a vanishing interval of time is 
arbitrarily large. Indeed, some proposals have been introduced to circumvent this unsuitable feature: refractory periods~\cite{18Evans,19MPuigdellosas} and intermittent potentials~\cite{20Mercado,21Santra}.  
The former considers an instantaneous reset that is followed by a residence time in the resetting state, whereas the latter considers a confining potential that randomly switches on/off in order to avoid the system to get away from the \textit{resetting state}. Notwithstanding, both models \textit{fails} someway. In a reset with refractory periods, the reset event itself is still instantaneous,
while intermittent potential does not guarantee the reset.

Return dynamics solves the problem~\cite{20Pal,19Pal,20Brodova,20Gupta,20Bressloff,21Zhou}. In stochastic returns, the dynamics comprises two evolution phases: the natural dynamics and the return dynamics. The first one is randomly interrupted
by switching it to the latter, which ends only when the resetting state is reached. Return phases have been considered to be either dynamically deterministic or stochastic. This strategy guarantees the effectiveness of realistic reset within a finite time. For the moment, experimentation has been scarce, although recent developments have allowed to start implementing stochastic resetting concepts in real experiments~\cite{20Tal,21Faisant}.

Stochastic thermodynamics~\cite{10sekimoto} addresses the study of thermodynamic quantities (work, heat, entropy, \dots) from the point of view of statistical mechanics, usually within a mesoscopic scale where fluctuations may be relevant. Surprisingly, to the best of our knowledge, there is not much literature on the stochastic resetting viewed under the lens of stochastic thermodynamics, apart a few exceptions~\cite{15Meylahn,16Fuchs,Deepak_PRL, Deepak_PRR, 17Pal}. Specifically, the cost in terms of mechanical work employed to implement the resetting dynamics have never been looked into deeply, despite its evident interest from both theoretical and experimental perspectives.
To fill in this gap, in this paper we aim to present a general method to compute the work fluctuations associated with stochastic return process facilitated by a confining potential.

Optimality have been addressed in the context of stochastic resetting~\cite{11Evans_b,13Evans}. Nevertheless, optimization problems have been posed traditionally for minimizing first passage time, usually motivated by applications to search processes. This is reasonable since no notion of cost was available before. Nevertheless, once the cost is introduced into this work, it is interesting to pose questions regarding optimization of such a cost or its fluctuations with respect to the resetting parameter.    

The rest of the articles is organized as follows. Section~\ref{sec:model} is devoted to the presentation of the model system that is a one-dimensional Brownian particle submitted to stochastic return. The central result of the article, which is the computation of the distribution of the mechanical work carried out the system, is derived in Sec.~\ref{sec:general}. The results are explicitly particularized in  Sec.~\ref{sec:results}. Therein, paradigmatic setups, Poissonian return either with V-shaped or harmonic potential, are used to illustrate the excellent agreement between theory and simulations.
Finally, we deliver conclusions and some perspectives in Sec.~\ref{sec:conclusions}. Some detailed calculations are relegated to the Appendix.

\section{The model}
\label{sec:model}
We consider a Brownian particle freely diffusing in a one-dimensional space with diffusion constant $D$. Thus, the time evolution of the particle position, $x(t)$, follows the Langevin equation:
\begin{equation}
    \dot x = \sqrt{2D}~\eta(t), \label{dyn-1}
\end{equation}
where the dot indicates a time-derivative, and  $\eta(t)$ is a Gaussian  white delta-correlated thermal noise with zero mean and unit variance, which fulfills $\langle \eta(t)\eta(t') \rangle = \delta(t-t')$. At a random interval of time, drawn from a specified distribution, $f(t)$, a confining potential, $U_r(x)$, with its minimum at the origin, is switched on. Once the potential is on, the dynamics reads
\begin{equation}
    \dot x = -\partial_x U_r(x) + \sqrt{2D}~\eta(t).\label{dyn-2}
\end{equation}
For the sake of simplicity, the same diffusion constant $D$ as in Eq.~\eref{dyn-1} has been considered in Eq.~\eref{dyn-2}. Nonetheless, results shown below can also be carried out for two different coefficients in the spirit of Ref.~\cite{20Gupta} along the same line.

The external potential is switched off when the system does the first passage to the origin, which is the minimum of the potential $U_r(x)$, and then, the system starts afresh the dynamics generated through Eq.~\eref{dyn-1} during a random time distributed according to $f(t)$. Then, dynamics in Eq.~\eref{dyn-2} starts up to the return to the origin and the same game is played iteratively. In summary, the system explores the space according to Eq.~\eref{dyn-1} ({\it exploration phase}), whereas it returns to the origin (the resetting location) following Eq.~\eref{dyn-2} ({\it return phase}). Notice that the above dynamics can be cast using only one Langevin equation:
\begin{equation}
\dot{x} = - \partial_x U(x, \lambda(t)) + \sqrt{2 D } \eta(t)
\end{equation}
for $U(x,\lambda(t)) \equiv \lambda(t) U_r(x)$, where  
$\lambda$ is a dichotomous controlled variable that switches between zero and unity for exploration and return phases, respectively. The stationary probability density function for the position emerging from this implementation of stochastic resetting has been studied in Ref.~\cite{20Gupta}, and the relaxation to the steady state is discussed in \cite{Gupta_2021_TD}.
In this article instead,  we focus on the thermodynamic properties of the system. Specifically, the distribution of the work, $W_{\rm tot}$, required to reset the system under the non-instantaneous resetting protocol discussed above up to a certain observation time $t$ is looked into.

On a general basis, within the framework of stochastic thermodynamics~\cite{10sekimoto},
the work in a dynamical process where the potential $U(x;\lambda(t))$ is varied through a control parameter $\lambda(t)$  can be simply defined, 
\begin{equation}
W_{\rm tot}(t)\equiv \int_0^t~{\rm d}t'~\partial_\lambda U(x;\lambda)~\dot \lambda. \label{work}
\end{equation}
For the model under consideration: (i) the aim of the external potential is to bring the system to the resetting location, which coincides with the minimum of the potential, (ii) contributions to the work are instantaneous since $\dot{\lambda}$ is different from zero just in a null set of times, which are those where the exploration phase is switched to the return phase or vice versa. For convenience, we consider $\min\{U(x;\lambda)\} = 0$, that has no physical implications since it just defines the energy origin.

As introduced above, the function $\lambda(t)$ is 
piece-wise constant. Let us define $t_i$ and $\tau_i$ as the times when, respectively, the $i$-th  exploration and return phase end.
Consistently, these times are stochastic. Specifically, the duration of the $i$-th exploration phase, $t_i-\tau_{i-1} $ ($\tau_0 \equiv 0$)  
has to be distributed according $f$;
whereas the duration of the $i$-th return phase $\tau_i-t_{i}$, follows 
a first passage distribution, which will be discussed later in detail. 
Hence, $\lambda(t)$ jumps from 0 to 1 at times $t_i$ and does the opposite at times $\tau_i$. Therefore, it is possible to explicitly write down:
\begin{equation}
\lambda(t) = \sum_{i=1}^{\infty} \Theta(t-t_i)-\sum_{j=1}^{\infty}\Theta(t-\tau_j),\label{lm-1}
\end{equation}
where $\Theta(\cdot)$ is the Heaviside theta function. 
The sum takes into account formally infinite events but if one is interested in a finite time window, it suffices to sum those within it. It is important to recall that the system starts in exploration phase, thus $0<t_1<\tau_1<t_2<\tau_2 < \cdots$.  

Differentiating  Eq.~\eref{lm-1}  with respect to time, and substituting $\dot \lambda$ into Eq.~\eref{work} yields
\begin{equation}
W_{\rm tot} (t)= \int_0^t~{\rm d}t'~U_r(x)
\bigg[\sum_{i=1}^{\infty} \delta(t'-t_i)-\sum_{j=1}^{\infty}\delta(t'-\tau_j)\bigg]~=\sum_{i=1}^{n_r(t)}  U_r(x(t_i)),
\label{work-2}
\end{equation}
where $n_r(t)$ counts the number of return phases started  
up to time $t$, and we have taken into account that $U_r(x(\tau_j))=0$ since $x(\tau_j)$ is the location of the minimum of the potential. Notice that $x(t_i)$, in Eq.~\eref{work-2}, is the position of particle where the resetting phase starts. Equation \eref{work-2} provides the work performed in a single trajectory. Of course, this is a stochastic quantity because of the underlying stochasticity of   the dynamics. The rest of this article is devoted to obtain the distribution of the work. 

\section{General theoretical framework}
\label{sec:general}
\begin{figure}
    \centering
    \includegraphics[width = 0.8 \textwidth]{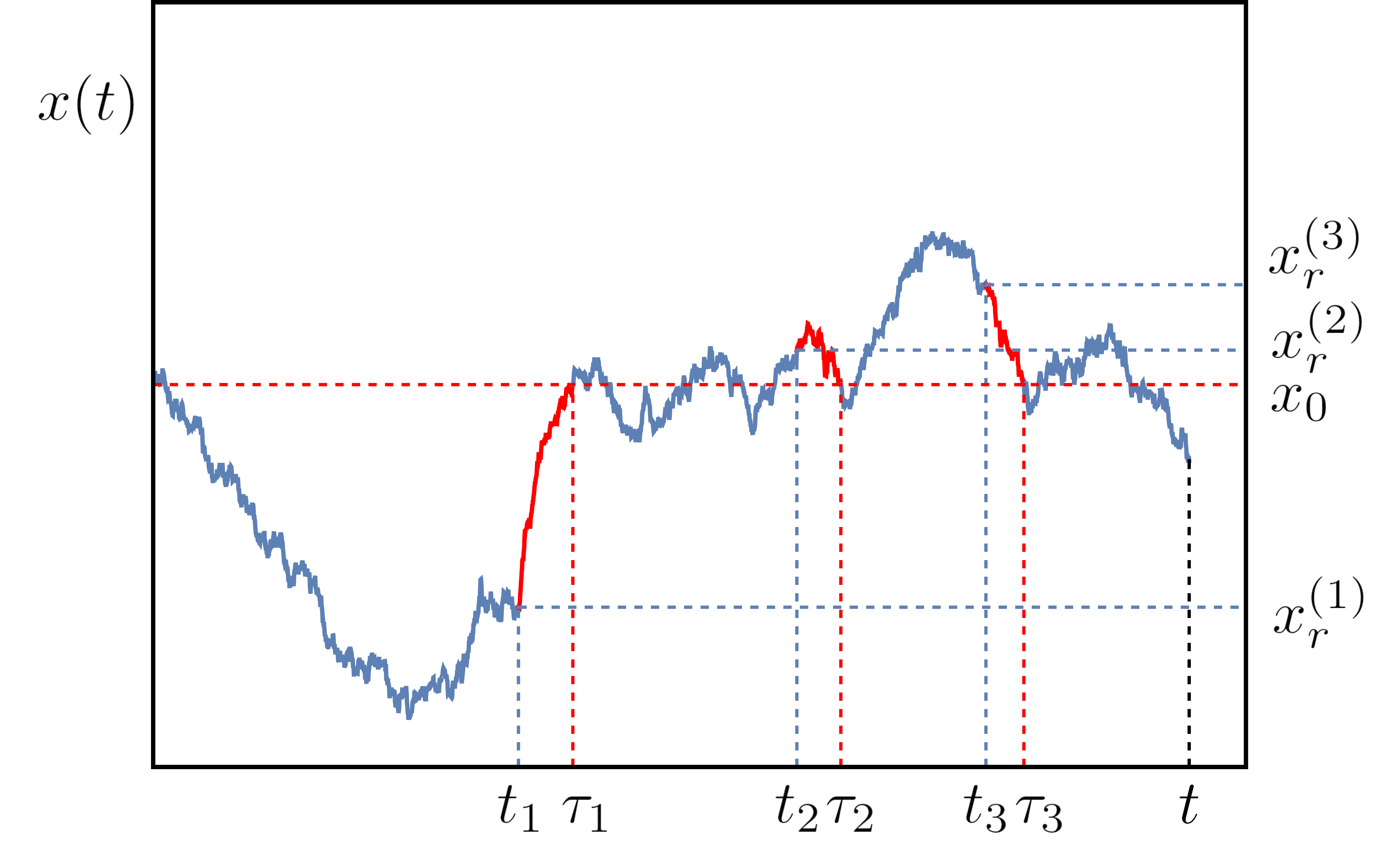}
    \caption{Sample trajectory of a Brownian particle submitted to stochastic return to location $x_0$. The evolution alternates between the free exploration phase (blue solid line) where the system follows its natural dynamics, and the return phase (red solid line), in which  a biased drift towards $x_0$ is introduced. The resetting pathway is defined by the set of times where a specific phase ends, either exploration, $t_i$ or return $\tau_i$, along with the positions where the return phases start, $x_r^{(i)}$. }
    \label{fig:res-path}
\end{figure}

Since the work in a single trajectory is written in terms of the position of the particle when return phases start, it will be required to analyze the statistics of what we call the {\it resetting pathway}. We define the resetting pathway as the specific resetting history followed by the Brownian particle, comprising the information concerning times where the phase is switched and position of the particle therein. Figure~\ref{fig:res-path} schematically illustrates the resetting pathway of a sample evolution. 

\subsection{Statistical weight of resetting pathway}
\label{conv-str}
This article aims at deriving the work distribution of the resetting system under study. The strategy to do so is to sum the contributions stemming from all possible resetting pathways  weighted consistently with their probability. For the sake of clarity, let us first introduce some useful notation:
\begin{enumerate}
\item $f(t)$ is the probability density function of the resetting time intervals.  
Defining $\tau_0 = 0$, $t_i-\tau_{i-1}$ is distributed according to $f$.
\item $F(t) \equiv \int_{t}^{\infty}~{\rm d}t' f(t')$ is the probability of not having started the return phase after  an exploration phase of duration $t$.
\item $p(x,t|x_0)$ is the probability density function of the position of a particle evolving an amount of time $t$ exclusively through the diffusion defined by the exploration phase given that it started at $x_0$. This is the solution of the Fokker-Planck equation associated to Eq.~\eref{dyn-1} with an initial condition $p(x,0|x_0)=\delta(x-x_0)$.
\item $\pi(x,\tau|x_r;x_0)$ is the probability density function of the position of a particle evolving an amount of time $\tau$ exclusively through the return phase given that it started at $x_r$ and there is an absorbing boundary at $x_0$. This is the solution of the Fokker-Planck equation associated to Eq.~\eref{dyn-2} in the returning phase with initial condition  $\pi(x,0|x_r;x_0)=\delta(x-x_r)$ with absorbing boundary condition  at $x_0$. To avoid cluttered formulae, in what follows we get rid of $x_0$ explicitly in the notation.  
For instance, this implies that for  $\pi$ and $p$, that is, we will  write simply $\pi(x,\tau|x_r)$ and $p(x,t)$ respectively. Note that we have chosen $x_0$ as the origin of coordinates indeed.  
\item $\phi(\tau|x_r)$ is the first passage time distribution to $x_0$ in the return phase given that the system starts from $x_r$. If a return phase starts at time $t_i$ from position $x_r$, the variable $\tau_i-t_i$ is distributed according to $\phi$.
\item $\Phi(\tau |x_r) \equiv 
\int_{\tau}^{\infty}~{\rm d}\tau' \phi(\tau'|x_r)$ is the probability of not having ended the return phase started a time $\tau$ ago, that is, the survival probability, $\Phi(\tau |x_r)=\int_{-\infty}^{\infty}~{\rm d}x~\pi(x,t|x_r)$ that fulfills $-\partial_\tau \Phi(\tau |x_r) = \phi(\tau)$. Note that $\pi$ is zero for $\textnormal{sign} (x-x_0) \neq \textnormal{sign} (x_r-x_0)$, since the absorbing boundary forbids the transfer of the particle to the other side of the boundary.  
\end{enumerate} 

The above statistical objects are the building blocks to characterize the statistical weight of a given resetting pathway. Let $\psi^{\rm p}_n(t)$ be the probability of having $n$ finished  trial  and being currently in the phase of  kind ${\rm p}=\{\textnormal{diff}, \textnormal{ret}\}$. (A finished trial means the completion of a return phase followed by an exploration phase.) Combining carefully the stay probabilities of each phase, and defining $\chi(x,t)\equiv 
f(t)p(x,t)$ for the sake of convenience, the general formulas for these probabilities read:
\begin{eqnarray}
\label{eq:psi-diff}
\psi^{\scriptsize \textnormal{diff}}_n(t)= &\left[  \prod_{i=1}^{n} \int_{\tau_{i-1}}^{t}\!\!\!\!\!{\rm d}t_i  \int_{-\infty}^{\infty} \!\!\!\!\!{\rm d}x^{(i)}_r \int_{t_{i}}^{t}\!\!{\rm d}\tau_i~ \chi \left(x^{(i)}_r,t_i-\tau_{i-1} \right) \phi \left( \tau_i-t_i|x^{(i)}_r \right) \right] \nonumber \\
&\times F(t-\tau_n) ,
\end{eqnarray}
and
\begin{eqnarray}
\label{eq:psi-ret}
\psi^{\scriptsize \textnormal{ret}}_n(t)=&\left[  \prod_{i=1}^{n} \int_{\tau_{i-1}}^{t}\!\!\!\!\!{ \rm d}t_i  \int_{-\infty}^{\infty} \!\!\!\!\!{\rm d}x^{(i)}_r \int_{t_{i}}^{t}\!\!{\rm d}\tau_i~ \chi \left(x^{(i)}_r,t_i-\tau_{i-1} \right) \phi \left( \tau_i-t_i|x^{(i)}_r \right) \right] \nonumber \\
& \times \int_{\tau_{n}}^{t}\!\!\!\!\!{\rm d}t_{n+1}  \int_{-\infty}^{\infty} \!\!\!\!\!{\rm d}x^{(n+1)}_r~\chi \left(x^{(n+1)}_r,t_{n+1}-\tau_{n}\right) \Phi \left( t-t_{n+1}|x^{(n+1)}_r \right) .
\end{eqnarray}
The integrand on the right-hand side of the above expressions are the probability density function for a specific resetting pathway. Of course, normalization
\begin{equation}
\label{eq:norm}
\sum_{n=0}^\infty \left[ \psi^{\scriptsize \textnormal{diff}}_n(t) + \psi^{\scriptsize \textnormal{ret}}_n(t)\right]=1,
\end{equation}
holds since the sum, plus all integrals inside the explicit form of $\psi$,
cover all possible resetting pathways. For a detailed discussion regarding the construction of the above probabilities as well as for a demonstration of its normalization,  see \ref{ap:prob}. 

Functions $\psi$ can be written in a compact form taking into account that they have a convolution structure. Using `$\ast$' to denote the convolution, i.e., $[A\ast B](t) \equiv \int_0^t~{\rm d}t'~A(t')B(t-t')$, between functions and as exponent, e.g., $A^{\ast 2} (t)= [A\ast A](t)$, for convolution power, it is possible to get 
\begin{equation}
\psi_n^{\scriptsize \textnormal{diff}} (t)= \left\{ \left[\int_{-\infty}^{\infty} {\rm d}x~\chi(x,\cdot)\ast\phi\left(\cdot|x\right)\right]^{\ast n}\ast F(\cdot) \right\} (t),
\end{equation} 
and
\begin{equation}
\psi_n^{\scriptsize \textnormal{ret}} (t)= \left\{ \left[\int_{-\infty}^{\infty} {\rm d}x~\chi(x,\cdot)\ast\phi\left(\cdot|x\right)\right]^{\ast n}\ast\left[\int_{-\infty}^{\infty} {\rm d}x~\chi(x,\cdot)\ast\Phi\left(\cdot|x\right)\right]\right\} (t).
\end{equation}

\subsection{Work distribution}
The convolution structure found out in the previous subsection
~\ref{conv-str} makes especially convenient to study the distribution of the work through the Laplace transform of its moment generating function. Let us start by defining the moment generating function:
\begin{equation}
\label{eq:def_GW}
G_W(k,t) \equiv   
\left\langle e^{k W_{\rm tot}(t)} \right\rangle,
\end{equation}  
where the notation $\langle \cdot \rangle$  stands for the weighted average over all possible resetting pathway. Since the total work is just the sum of contributions of $U_r$ evaluated at the particle's positions where the return phases start, we have that 
\begin{eqnarray}
G_W(k,t) = &\sum_{n=0}^{\infty} \left\{ \left[\int_{-\infty}^{\infty} {\rm d}x~ \chi_W(k,x,\cdot)\ast\phi\left(\cdot|x \right)  \right]^{\ast n}\ast F(\cdot)\right\} (t)	\nonumber \\
& +\sum_{n=0}^{\infty} \left\{ \left[\int_{-\infty}^{\infty} {\rm d}x~\chi_W(k,x,\cdot)\ast\phi\left(\cdot|x\right) \right]^{\ast n} \right. \nonumber \\
& \qquad \qquad  \left. \ast\left[\int_{-\infty}^{\infty} {\rm d}x~\chi_W(k,x,\cdot)\ast\Phi\left(\cdot|x\right)\right]\right\} (t)
\end{eqnarray} 
with $\chi_{W}(k,x,t)  \equiv   
\chi(x,t) e^{k U_r(x)}.$

Introducing the Laplace transform,
\begin{equation}
\widetilde{g}(s)  \equiv 
\int_0^{\infty}{\rm dt}~ e^{-st} g(t),
\end{equation}
we can write explicitly the Laplace transform of the moment generating functions
\begin{equation}
\label{eq:Lap_gen_W}
\widetilde{G}_W(k,s) = \frac{\widetilde{F}(s)+\frac{1}{s}\int_{-\infty}^{\infty}{\rm d}x~\widetilde{\chi}_W(k,x,s)\left[1 - \widetilde{\phi}(s|x) \right]}{1-\int_{-\infty}^{\infty}{\rm d}x~\widetilde{\chi}_W(k,x,s) \widetilde{\phi}(s|x) }
\end{equation}
in terms of the Laplace transforms of the functions characterizing the probabilistic ingredients of the model. Above, we have used the convolution theorem for Laplace transform,  
carried out the sum of the geometric series and recalled the relation $\Phi(\tau|x_r) = \int_{\tau}^{\infty}d\tau' \phi(\tau'|x_r)$. 

Equation \eref{eq:Lap_gen_W} is the central result of this study. It is an exact general result for the distribution work of a system submitted to return dynamics with arbitrary return potential $U_r$ and arbitrary return rate $f$. 

\section{Results for diffusion under Poissonian return with paradigmatic potentials}
\label{sec:results} 
From now on, we consider the case of Poissonian return, $f(t)= 
re^{-rt}$, with $r$ being a constant resetting rate. This choice allows to give some explicit simplifications in Eq.~\eref{eq:Lap_gen_W}. Specifically, we get
\begin{equation}
\label{eq:Lap_gen_W_pois}
\widetilde{G}_W(k,s) = \frac{\frac{1}{s+r}+\frac{r}{s}\int_{-\infty}^{\infty}{\rm d}x~e^{k U_r(x)} \widetilde{p}(x,s+r) \left[1 - \widetilde{\phi}(s|x) \right]}{1-r\int_{-\infty}^{\infty}{\rm d}x~e^{k U_r(x)} \widetilde{p}(x,s+r)  \widetilde{\phi}(s|x) },
\end{equation}
where 
\begin{equation}
\label{eq:p-diff-lt}
\widetilde{p}(x,s) = \frac{e^{-\sqrt{\frac{s}{D}}|x|}}{\sqrt{4Ds}}. 
\end{equation}
is the Laplace transform of the free propagator of the diffusion phase, $p(x,t)=
\exp\left[-x^2/(4Dt)\right]/\sqrt{4\pi D t}$.

In the following, we consider two specific paradigmatic potentials: V-shaped and harmonic.

\subsection{V-shaped potential}
\label{mod-pot}
We consider first a V-shaped potential, $U_r(x) = 
\gamma_V |x|$ in the return phase. 
This is one of the quite few cases, where the first passage density, $\phi(t|x)$, for the particle to reach the origin starting from location $x$ can be obtained exactly, see section 3.2.2.2 from Ref.~\cite{01Redner_book},
\begin{equation}
\phi(t|x)=\frac{|x|}{\sqrt{4 \pi D t^{3}}} \exp\bigg[-\frac{(|x|-\gamma_V  t)^2}{4 D t}\bigg],
\label{fpd}
\end{equation}
and its Laplace transform reads
\begin{equation}
\widetilde{\phi}(s|x)=\exp\left[\frac{\gamma_V |x|}{2D}\bigg(1-\sqrt{1+\frac{4 D~s}{\gamma_V^2}}\bigg)\right]. \label{fp-lt}
\end{equation}

Substituting Eqs.~\eref{eq:p-diff-lt} and \eref{fp-lt} into Eq.~\eref{eq:Lap_gen_W_pois} and performing the integrals over $x$,
\begin{equation}
    \widetilde{G}_W(k,s) = \frac{\frac{1}{r+s}+\frac{r \left(\sqrt{\alpha_V +s}-\sqrt{\alpha_V }\right)}{s \sqrt{r+s} \left(\sqrt{r+s}-2 D k \sqrt{\alpha_V }\right) \left[\sqrt{r+s}+\sqrt{\alpha_V +s}-\sqrt{\alpha_V } (2 D k+1)\right]}}{1-\frac{r}{\sqrt{r+s} \left[\sqrt{r+s}+\sqrt{\alpha_V +s}-\sqrt{\alpha_V } (2 D k+1)\right]}}
    \label{eq:G_V_exact}
\end{equation}
where, for the sake of simplicity, we have introduced  the parameter  $\alpha_V \equiv 
\gamma_V^2 /4D$, which characterizes the inverse of a time-scale, defined by the relative intensity of the confining potential with respect to the diffusion. 

Moments can be obtained from the moment generating function~\eref{eq:Lap_gen_W}. Specifically the  $n$-th moment of the work in the Laplace space is
\begin{equation}
   \widetilde{\langle W^n\rangle}(s) \equiv \frac{\partial^n}{\partial k^n} \widetilde{G}_W(k,s)\bigg|_{k=0}. \label{def-mom}
\end{equation}
We have carried out the explicit calculation for the first and second moments.
 On the one hand, the expression obtained exactly are not especially illuminating and then we relegate them to \ref{ap:V-exact}. On the other hand, the limit of large times is quite informative and fairly simple. Thus, we develop it here. To look into large times, it is needed to study the Laplace transform for small $s$. Specifically, the Laurent series for the two first moments read:
\begin{eqnarray}
\widetilde{\langle W\rangle}(s) &=&  \mu_{1}^{(-2)} (D,r,\alpha_V)\frac{1}{s^2} + \mu_1^{(-1)}(D,r,\alpha_V) \frac{1}{s} + \mathcal{O}\left(s^{0}\right),\label{ILT-mu1} \\ %
\widetilde{\langle W^2\rangle}(s) &=& \mu_{2}^{(-3)}(D,R,\alpha_V) \frac{1}{s^3} + \mu_{2}^{(-2)}(D,R,\alpha_V)\frac{1}{s^2} + \mathcal{O}\left(s^{-1}\right). 
\end{eqnarray}
with
\begin{eqnarray} 
\mu_{1}^{(-2)}(D,r,\alpha_V) \equiv \frac{4   D \sqrt{r} \alpha_V}{\sqrt{r}+2 \sqrt{\alpha_V }},  \\ 
\mu_1^{(-1)}(D,r,\alpha_V) \equiv \frac{D\left(r^{3/2}+2r \sqrt{\alpha_V }-4 \alpha_V ^{3/2}\right)}{\sqrt{r} \left(\sqrt{r}+2 \sqrt{\alpha_V }\right)^2},\\
\mu_{2}^{(-3)}(D,r,\alpha_V) \equiv \frac{32  D^2 r\alpha_V ^2}{\left(\sqrt{r}+2 \sqrt{\alpha_V }\right)^2}, \\
 \mu_{2}^{(-2)}(D,R,\alpha_V) \equiv \frac{16   D^2 r \alpha_V \left(\sqrt{\alpha_V }+\sqrt{r}\right)}{\left(\sqrt{r}+2 \sqrt{\alpha_V }\right)^3}.
\end{eqnarray}

Application of the generalized final value theorem allows to study the asymptotic behavior of the first moments. The mean work diverges linearly with time, 
\begin{equation}
\lim_{t \to \infty} \frac{\langle W \rangle (t)}{t} = \mu_1^{(-2)} (D,r,\alpha_V)= \frac{4   D \sqrt{r} \alpha_V}{\sqrt{r}+2 \sqrt{\alpha_V }},
\label{eq:mean}
\end{equation}
whereas the second moment does it quadratically
\begin{equation}
\lim_{t \to \infty} \frac{\langle W^2 \rangle (t)}{t^2} = \frac{\mu_2^{(-3)} (D,r,\alpha_V)}{2}=\frac{16  D^2 r\alpha_V ^2}{\left(\sqrt{r}+2 \sqrt{\alpha_V }\right)^2}.
\end{equation}
Nevertheless, the variance of the work, $\sigma_W^2 \equiv \langle W^2 \rangle (t) - \langle W \rangle^2 (t)$, behaves linearly in the  long-time regime, 
\begin{eqnarray}
\lim_{t \to \infty} \frac{\sigma_W^2 (t)}{t}& = \mu_2^{(-2)} (D,r,\alpha_V)-2 \mu_1^{(-2)} (D,r,\alpha_V)\mu_1^{(-1)} (D,r,\alpha_V) \nonumber\\
&=\frac{8 \alpha_V  D^2 \left(r^{3/2}+4 \alpha_V ^{3/2}\right)}{\left(\sqrt{r}+2 \sqrt{\alpha_V}\right)^3},
\label{eq:var}
\end{eqnarray}
since $\left[\mu_1^{(-2)}\right]^2=\mu_2^{(-3)}/2$.  Provided the asymptotic behavior for the mean and the variance, one trivially gets that, for the square of the coefficient of variation,
\begin{equation}
\lim_{t\to \infty} \frac{t\sigma_W^2(t)}{\langle W \rangle^2(t)}= \frac{r^{3/2}+4\alpha_V^{3/2}}{2 r \alpha_V (\sqrt{r}+2 \sqrt{\alpha_V})}.
\label{eq:CV}
\end{equation}
Off course, the coefficient of variation without the $t$-rescaling, goes to zero for large times. Thus, the relative fluctuations being negligible in such regime. 

\begin{figure}
    \centering
    \includegraphics[width = \textwidth]{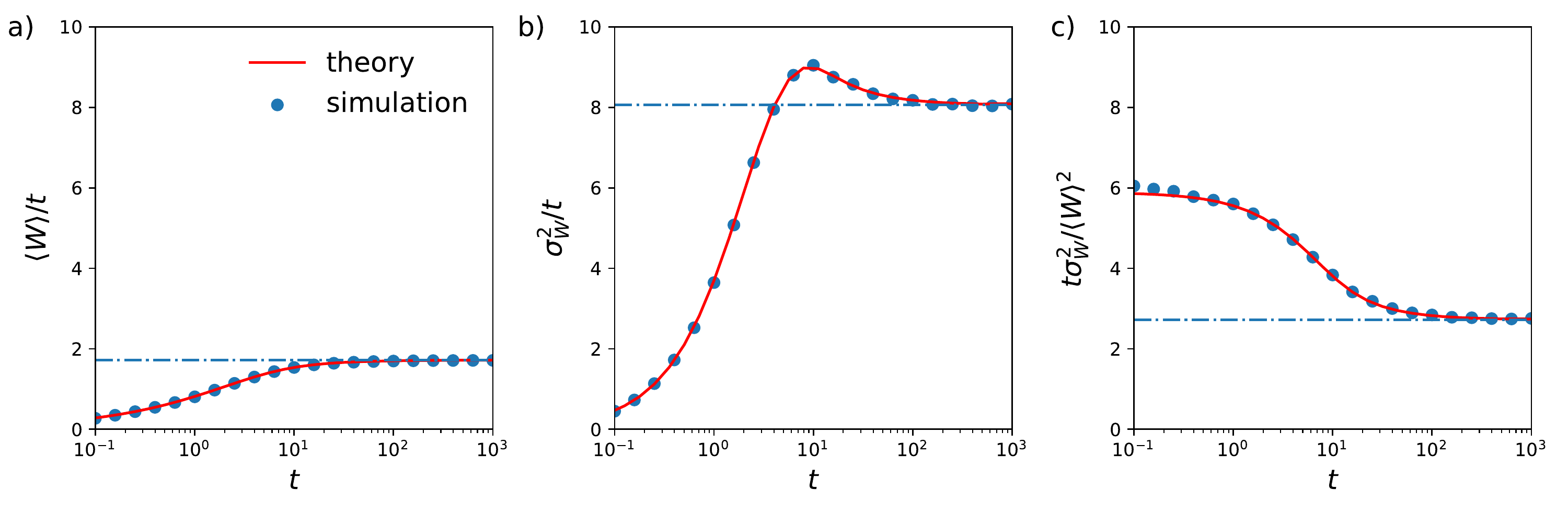}
    \caption{Rescaled mean (a), variance (b), and the square of the coefficient of variation (c) for a particle submitted to V-shaped potential under return dynamics. Theoretical predictions for the time evolution stems from numerical inversion of the Laplace transform of the moments (see \ref{ap:V-exact} for their analytical forms) whereas the asymptotic values comes from Eqs.~\eref{eq:mean}, \eref{eq:var} and \eref{eq:CV}.  The parameters considered here are $r = 0.3$, $D = 2$, and $\alpha_V=1$. Simulation results stand for average over $10^5$ trajectories using a time step $dt = 10^{-3}$. }
    \label{fig:cumulants}
\end{figure}

The theoretical predictions derived for the V-shaped potential are compared to results obtained from averaging over simulated trajectories in Fig.~~\ref{fig:cumulants}. Therein, we have obtained the theoretical evolution of the moments by numerical inversion of the Laplace transform of the moments, whose analytical expression can be found in\ref{ap:V-exact}. The prediction of the asymptotic value for large times is obtained directly from Eqs.~\eref{eq:mean}, \eref{eq:var} and \eref{eq:CV}. The agreement is excellent, not only for the asymptotic regime but also in the time evolution predicted by the numerical inversion.

\begin{figure} 
    \centering
    \includegraphics[width = 0.49\textwidth]{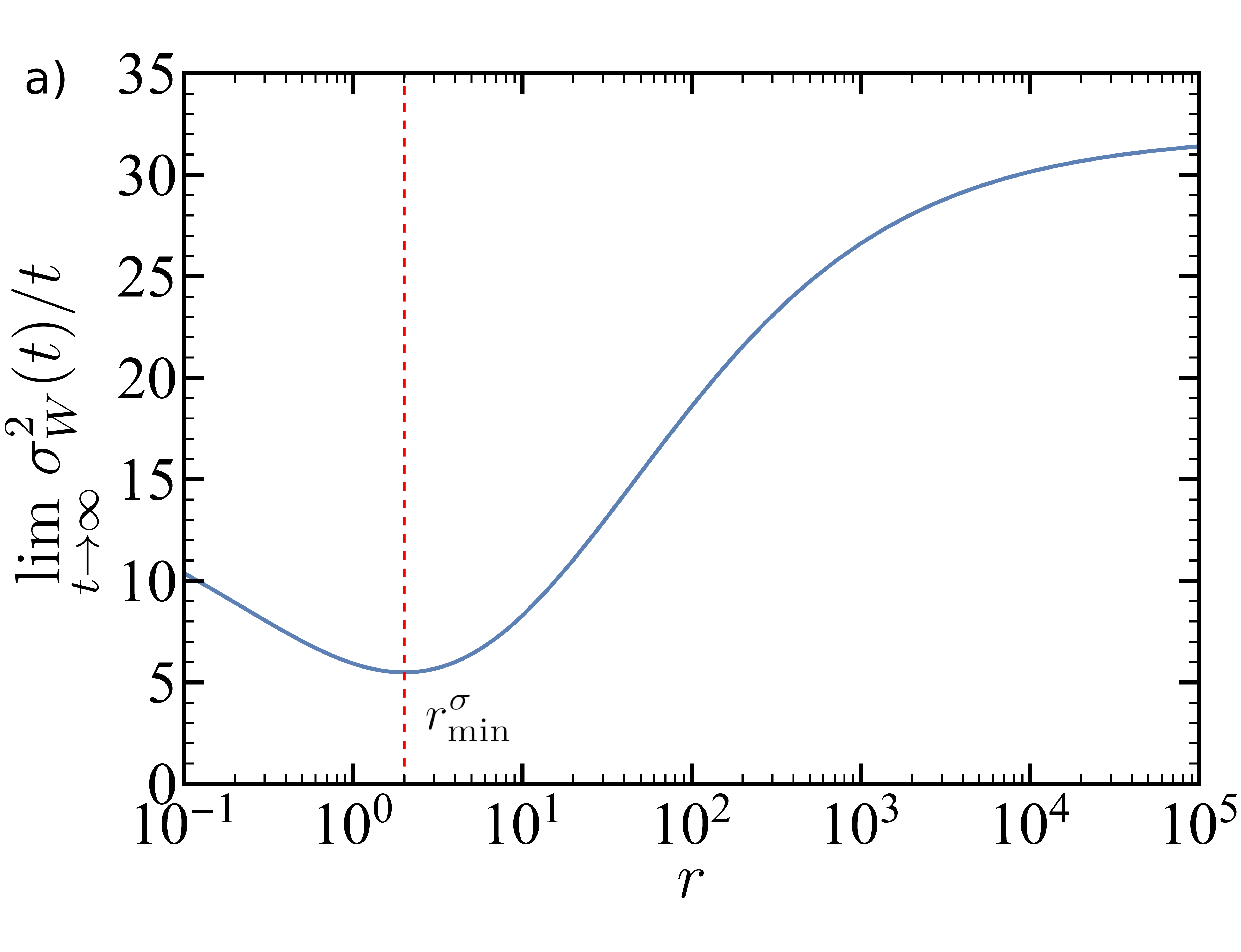}
    \includegraphics[width = 0.49\textwidth]{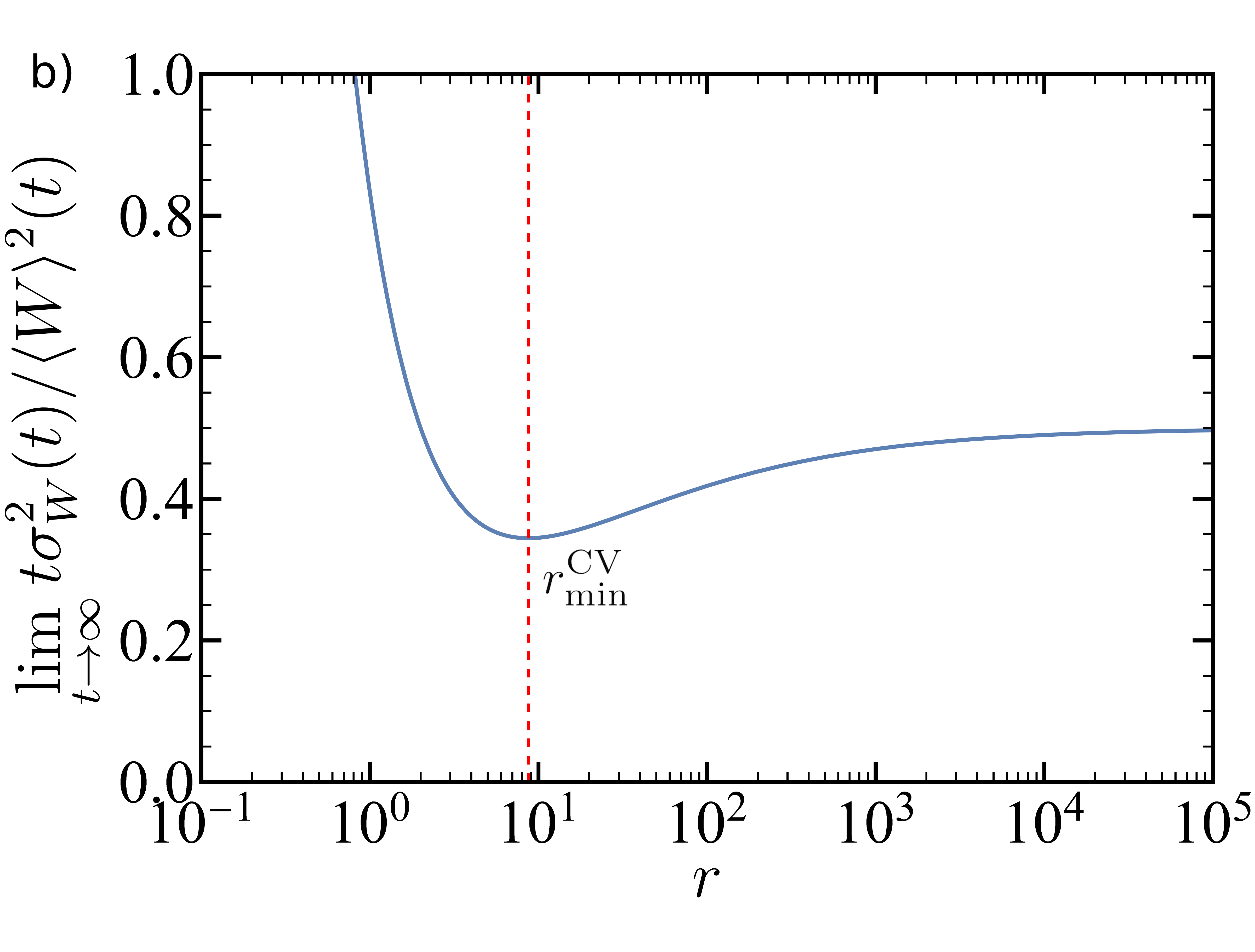}
    \caption{Asymptotic rescaled fluctuations as a function of the resetting rate. Variance (a) and the square of the coefficient of variation (b) are non-monotonous functions as given by Eqs.~\eref{eq:var} and \eref{eq:CV} respectively. The vertical red dashed lines stand for the optimal values minimizing the fluctuations in Eqs.~\eref{eq:rminvar} and \eref{eq:rmincv}. As in Fig.~\ref{fig:cumulants}, $D=2$ and $\alpha_V=1$. Regardless the parameters, the optimal resetting rate minimizing the coefficient of variation is roughly four times the one minimizing the variance.}
    \label{fig:opt}
\end{figure}

Let us discuss in detail the dependence of the asymptotic values obtained above on the return rate $r$. During the whole discussion, we make reference to the $t-$rescaled moments appearing in Eqs.~\eref{eq:mean}, \eref{eq:var} and \eref{eq:CV}. The rescaled mean work is an increasing function of $r$ that goes from $\langle W \rangle/t= 0$ for $r=0$, and possesses a horizontal asymptote at $\langle W \rangle/t = 4 D \alpha_V$ for $r \to \infty$. Both, the variance and the coefficient of variation are non-monotonous functions of $r$. Both of them start decreasing from $r=0$ up to find a minimum value and then increase up to reach a horizontal asymptote (see Fig.~\ref{fig:opt})
The values of the return rate $r_{\min}$ at which the minimum is attained are 
\begin{eqnarray}
\label{eq:rminvar}r^{\sigma}_{\min}= 2 \alpha_V, \\
\label{eq:rmincv}r^{\rm CV}_{\min}=2(3-2\sqrt{2})^{1/3}[1+(3+2\sqrt{2})^{1/3}]^2 \alpha_V \simeq 8.71 \alpha_V,
\end{eqnarray} 
for $\sigma_W^2/t$ and $t \sigma_W^2/\langle W \rangle^2$, respectively. 
This is a remarkable feature from the optimization point of view. The dependence of the work fluctuations on the return rate are non-trivial, exhibiting a minimum value for the aforementioned values either in absolute or relative terms. 

\subsection{Harmonic potential}

We consider now a harmonic potential, $U_r(x) = \gamma_h x^2/2$ in the return phase. 
This is another simple case, where the first passage density, $\phi(t|x)$, for the particle to reach the origin starting from location $x$ can be obtained (see for instance Appendix A.2 in Ref.~\cite{20Gupta}). The result in the Laplace domain is 
\begin{equation}
\widetilde{\phi}(s|x)=\sqrt{\frac{\gamma_h x^2}{2 \pi D }}~\Gamma \left(\frac{s+\gamma_h}{2 \gamma_h}\right)~\mathcal{U}\left(\frac{s+\gamma_h}{2\gamma_h},\frac{3}{2},\frac{\gamma_h x^2}{2 D}\right),\label{fpd-ho} 
\label{fpd}
\end{equation}
where $\Gamma$ and $\mathcal{U}$ stand for the gamma function and the confluent hypergeometric function, also known as Tricomi's function, respectively.
Obtaining the Laplace transform of the moment generating function for the work reduces to substitute Eqs.~\eref{eq:p-diff-lt} and \eref{fpd} into Eq.~\eref{eq:Lap_gen_W_pois} with $U_r(x)=\gamma_h x^2/2$ and performing the integrals over $x$.

Unlike the case of V-shaped potential discussed in the previous subsection~\ref{mod-pot}, the analytical computation of the first and second moment is not straightforward. Nevertheless, it is possible to compute these quantities in the long-time limit numerically. Let us briefly sketch the procedure for the computation of the scaled moments. For a given Laplace transform variable $s$, we compute the first and second moment (in the Laplace space), respectively, by taking first and second order derivative of the numerical expression $\widetilde{G}_W(k,s)$ with respect to $k$ and set those derivatives equal to zero [see Eq.~\eref{def-mom}]. By numerically inverting the Laplace transform, we can compute the first and second scaled cumulant in the long-time limit. 

\begin{figure}
    \centering
    \includegraphics[width = 15cm]{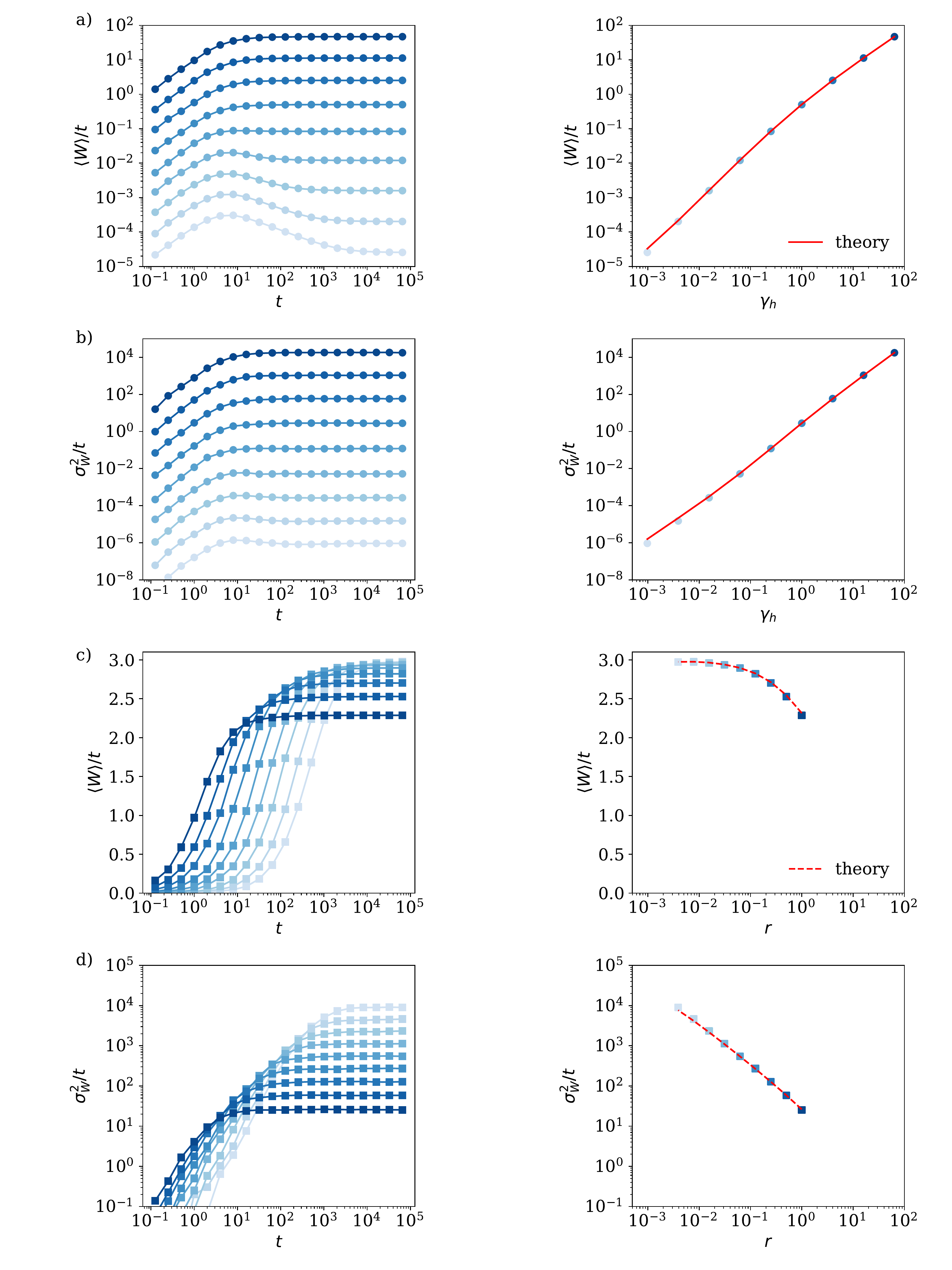}
    \caption{First and second scaled cumulant of work for stochastic returns facilitated by a harmonic trap. Left panel (a-d): Figures are obtained using Langevin simulations. Right panel (a-d): Figures show the plateau points of each figure in the left panel, and show the comparison of analytical results with Langevin simulation in the long-time time. (a-b): Resetting rate $r=0.5$. (c-d): Trap stiffness $\gamma_h=0.5$. In all plots, the diffusion constant $D=0.75$ is kept fixed. The numerical simulation is performed for a time increment $dt=10^{-3}$, and the averaging is performed over $10^4$ realizations. }
    \label{ho-fig}
\end{figure}

In Fig.~\ref{ho-fig}(a-d), we show the comparison of these scaled cumulants obtained using Langevin simulations for different sets of parameters. Furthermore, we show the comparison of the long-time analytical results with long-time numerical simulations results.
Clearly, we see that the scaled cumulants becomes independent of time similar to what we have observed in the previous subsection~\ref{mod-pot} for the case of V-shaped potential.  Further, we notice from the right panel that the scaled cumulants of the work increase (decreases) with the stiffness parameter, $\lambda$ (resetting rate, $r$) in the long-time limit. 

\section{Conclusions}
\label{sec:conclusions}

As far as we are aware, this is the first time that the distribution of the mechanical work performed by a resetting mechanism is worked out. Our framework, relying on the average over resetting pathways, leads to write the distribution---the Laplace transform of the moment generating function to be more accurate---in terms of statistical information of the processes that constitute the return dynamics, see Eq.~\eref{eq:Lap_gen_W}. Note that this approach remains completely valid for both arbitrary resetting time distributions $f(t)$ and arbitrary dynamics for the diffusion and return phases. Furthermore, not only is the framework useful for computing the work distribution but any physical observable depending on the resetting pathway. 

The power to obtain exact results through this method has been shown through explicit computation in paradigmatic cases. Specifically, a V-shaped potential, which confine particles by submitting to constant force to both sides of the confining point, and harmonic potential have 
been considered under Poissonian resetting times.

For the first one, explicit calculations are manageable, being possible to reach a closed expression for the Laplace transform of the moment generating function. Hence, Laplace transform of the moments is obtained exactly and validated with simulation results pleasurably after inverting numerically the Laplace transform. The asymptotic behavior for large times is fully resolved analytically for the first moments. An optimal resetting rate emerges when analyzing the fluctuations of the work.  This is a remarkable feature characterizing the non-trivial behavior of the work.

Regarding the harmonic case, the explicit calculation is not so straightforward as for the V-shaped one. Nevertheless, numerical evaluation of the predictions and comparison with simulation results have validated our approach in this more involved case.

The methods introduced here are expected to open new insight to stochastic thermodynamics in resetting systems, allowing to look into the energetics of the resetting mechanism and its cost. Optimal features, as the one shown for the V-shaped potential, are especially appealing since they are the hallmark for engineering efficient resetting mechanisms.

\section*{Acknowledgments}
The authors thank Prof.~Anupam Kundu (ICTS India) for many fruitful discussions.
C.A.P. acknowledges financial support from Grant PGC2018-093998-B-I00 funded by MCIN/AEI/10.13039/501100011033/ and by ERDF ``A way of making Europe.'' as well as the funding from Junta de Andaluc\'{\i}a and European Social Fund through the program PAIDI-DOCTOR.

\appendix

\section{Detailed derivation of the probability of observing a certain phase}
\label{ap:prob}

Herein, we build the probability of observing our system at time $t$ within a certain phase, either exploration or return, after a given number of finished/successful trials. (Again a finished trial means the completion of a return phase followed by an exploration phase.) This is a generalization of the framework introduced in the section IA of the supplementary material of Ref.~\cite{Deepak_PRL}, where analogous procedure was carried out for the simpler case of instantaneous resetting. 

We recall that $\psi^{{\rm p}}_n(t)$ stands for the probability at time $t$ of having $n$ finished trials,
and being currently in the phase of kind ${\rm p }=\{{\rm diff},{\rm ret}\}$. Our strategy here is  writing explicitly the first probabilities $\psi^{{\rm p}}_n(t)$ and then generalize by induction. The very first one is  $\psi^{\rm diff}_0(t)$, which is simply the probability of not having started the first return up to time $t$, that is,
\begin{equation}
\psi^{\rm diff}_0(t)= F(t).
\end{equation}
The probability of having first reset and not having finished the first return phase is
\begin{equation}
\psi^{\rm ret}_0(t)= \int_0^{t}\!\!{\rm d}t_1 \int_{-\infty}^{\infty} \!\!\!\!\!{\rm d}x^{(1)}_r  f(t_1) p\left(x^{(1)}_r,t_1\right) \Phi \left( t-t_1|x^{(1)}_r \right),
\end{equation}
where we have taken into account that the first return starts at time $t_1$, from $x_r^{(1)}$, but this return phase has not finished at time $t$ yet. Hence, it is clear why, in the main text, we have conveniently introduced the function $\chi(x,t)=f(t)p(x,t)$ [see Eqs. \eref{eq:psi-diff} and \eref{eq:psi-ret}]. Of course, to obtain the probability $\psi_0^{\rm ret}$, the integrals over the intermediate time and position $t_1$ and $x_r^{(1)}$ have to be carried out.   
We can go further and write the probability of having finished the first trial
but not having finished the subsequent exploration phase
 \begin{eqnarray}
 \psi^{\rm diff}_1(t)=&\int_0^{t}\!\!{\rm d}t_1 \int_{-\infty}^{\infty} \!\!\!\!\!{\rm d}x^{(1)}_r \int_{t_1}^{t} \!\! {\rm d}\tau_1 f(t_1) p\left(x^{(1)}_r,t_1\right) \phi \left( \tau_1-t_1|x^{(1)}_r \right) \nonumber \\
& \times F(t-\tau_1).
 \end{eqnarray}
 Above, we have a first exploration phase finishing at $t_1$, when the return phase starts from $x_r^{(1)}$, reaching the resetting point $x_0$ at time $\tau_1$. Then a subsequent exploration phase starts without finishing. Next, we write the probability of first finished trial with the second exploration phase ends at time $t_2$, but the second return phase does not end until the observation time $t$:
 \begin{eqnarray}
      \psi^{\rm ret}_1(t)=&\int_0^{t}\!\!{\rm d}t_1 \int_{-\infty}^{\infty} \!\!\!\!\!{\rm d}x^{(1)}_r \int_{t_1}^{t} \!\! {\rm d}\tau_1 \int_{\tau_1}^{t}\!\!{\rm d}t_2 \int_{-\infty}^{\infty} \!\!\!\!\!{\rm d}x^{(2)}_r f(t_1) p\left(x^{(1)}_r,t_1\right) \nonumber \\
 &\times  \phi \left( \tau_1-t_1|x^{(1)}_r \right) f(t_2-\tau_1) p\left(x^{(2)}_r,t_2-\tau_1 \right) 
 \nonumber \\
 &\times \Phi\left( t-t_2|x^{(2)}_r \right).
 \end{eqnarray}
The iterative construction of more and more complex resetting pathways is straightforward, implying the addition of extra terms in the convolution structure. Specifically, when we pass from $n$ to $n+1$, the product $f p \phi$ with the proper arguments is added to the convolution, representing a full trial (exploration $+$ return) of the dynamics. This property is the fingerprint of renewal structure. The general expression for $\psi^{\rm p}_n$ is provided in the main text in Eqs.~\eref{eq:psi-diff} and \eref{eq:psi-ret}. 

Note that considering instantaneous resetting, $\phi(\tau|x_r)=\delta(\tau)$ and $\Phi(\tau|x_r)=0$, thus, the results in Ref.~\cite{Deepak_PRL} are reobtained.

The normalization property \eref{eq:norm} can be easily demonstrated. Specifically, the sum of probabilities can be written as the generating function \eref{eq:def_GW} evaluated at $k=0$. Its Laplace transform is
\begin{equation}
\widetilde{G}(0,s) = \frac{\widetilde{F}(s)+\frac{1}{s}\widetilde{f}(s)-\frac{1}{s}\int_{-\infty}^{\infty}{\rm d}x~\widetilde{\chi}(x,s) \widetilde{\phi}(s|x) }{1-\int_{-\infty}^{\infty}{\rm d}x~\widetilde{\chi}(x,s) \widetilde{\phi}(s|x) }
\end{equation}
where we have rewritten Eq.~\eref{eq:Lap_gen_W} using
explicitly the function $\chi(x,t)=f(t)p(x,t)$ in the numerator and $\int_{-\infty}^{\infty} {\rm d} x_r \, p(x_r,t)=1$. Finally, using the relation  $F(t)=\int_t^{\infty} {\rm d}t' \, f(t')$, that implies $\widetilde{F}(s)=[1-\widetilde{f}(s)]/s$, we get
\begin{equation}
\widetilde{G}(0,s)= \frac{1}{s} \Rightarrow \widetilde{G}(0,t)=1,
\end{equation}
which ensures the normalization of resetting pathways, as expected.

\section{Exact expressions for the first and second moments of the work in the V-shaped potential}
\label{ap:V-exact}

We provide below the exact expressions for the first and second moments of the work in the Laplace space for the Brownian particle submitted to stochastic return driven by a V-shaped potential. The results are obtained by direct evaluation of Eq.~\eref{def-mom} using Eq.~\eref{eq:G_V_exact}.

For the first moment, we get
\begin{eqnarray}
\langle \widetilde{W} \rangle(s) =  &2 \sqrt{\alpha_V } D r \Bigg[(2 \alpha_V  \sqrt{r+s}-2 \sqrt{\alpha_V +s} \sqrt{\alpha_V  (r+s)} \nonumber \\
&+r \left(\sqrt{\alpha_V +s}-\sqrt{\alpha_V }\right)+ 2 s \left(\sqrt{r+s}+\sqrt{\alpha_V +s}-\sqrt{\alpha_V }\right) \Bigg] \nonumber \\ & \times s^{-1} (r+s)^{-1}   \left(s+\sqrt{r+s} \sqrt{\alpha_V +s}-\sqrt{\alpha_V  (r+s)}\right)^{-2}. \label{mu-1}
\end{eqnarray}

For the second moment, it is possible to obtain
\begin{equation}
 \langle \widetilde{W^2} \rangle (s) = \frac{8  D^2 r \alpha_V}{\zeta_4} \left[r^3 \left(\sqrt{\alpha_V +s}-\sqrt{\alpha_V }\right)+2 r^2 \zeta_1+2 r \zeta_2+4 s \zeta_3\right], \label{mu-2},
\end{equation}
where
\begin{eqnarray}
    \zeta_1\equiv&-8 \alpha_V ^{3/2}+3 \alpha_V  \sqrt{r+s}-3 \sqrt{\alpha_V +s} \sqrt{\alpha_V  (r+s)} \nonumber \\
    & +s \left(-9 \sqrt{\alpha_V }+2 \sqrt{r+s}+5 \sqrt{\alpha_V +s}\right)+8 \alpha_V  \sqrt{\alpha_V +s},\\
    \zeta_2  \equiv & s^2 \left(-27 \sqrt{\alpha_V }+8 \sqrt{r+s}+12 \sqrt{\alpha_V +s}\right) \nonumber \\
    & + 4 \alpha_V ^{3/2} \left(\sqrt{\alpha_V }-\sqrt{\alpha_V +s}\right) \left(3 \sqrt{r+s}-2 \sqrt{\alpha_V }\right) \nonumber \\ 
    &+s \bigg(-36 \alpha_V ^{3/2}+23 \alpha_V  \sqrt{r+s} 
    \nonumber \\
    & \qquad -17 \sqrt{\alpha_V +s} \sqrt{\alpha_V  (r+s)}+32 \alpha_V  \sqrt{\alpha_V +s}\bigg),\\
    \zeta_3  \equiv&2 s^2 \left(2 \left(\sqrt{r+s}+\sqrt{\alpha_V +s}\right)-5 \sqrt{\alpha_V }\right) \nonumber \\
    &+2 \alpha_V ^{3/2} \left(\sqrt{\alpha_V }-\sqrt{\alpha_V +s}\right) \left(5 \sqrt{r+s}-2 \sqrt{\alpha_V }\right)\nonumber\\
    &+s \bigg(-15 \alpha_V ^{3/2}+15 \alpha_V  \sqrt{r+s}-10 \sqrt{\alpha_V +s} \sqrt{\alpha_V  (r+s)} \nonumber \\
    & \qquad +13 \alpha_V  \sqrt{\alpha_V +s}\bigg),\\
    \zeta_4  \equiv& s (r+s)^{3/2} \left(-\sqrt{\alpha_V }+\sqrt{r+s}+\sqrt{\alpha_V +s}\right)^2 \nonumber\\
    & \times \left(-\sqrt{\alpha_V  (r+s)}+\sqrt{r+s} \sqrt{\alpha_V +s}+s\right)^3.
\end{eqnarray}

\section*{References}

\begin{thebibliography}{10}

\bibitem{11Evans}
Martin~R. Evans and Satya~N. Majumdar.
\newblock Diffusion with stochastic resetting.
\newblock {\em Phys. Rev. Lett.}, 106:160601, Apr 2011.

\bibitem{Pal_PRE}
Arnab Pal.
\newblock Diffusion in a potential landscape with stochastic resetting.
\newblock {\em Phys. Rev. E}, 91:012113, Jan 2015.

\bibitem{Sanjib_relax}
Satya~N. Majumdar, Sanjib Sabhapandit, and Gr\'egory Schehr.
\newblock Dynamical transition in the temporal relaxation of stochastic
  processes under resetting.
\newblock {\em Phys. Rev. E}, 91:052131, May 2015.

\bibitem{16Mendez}
Vicen\ifmmode \mbox{\c{c}}\else~\c{c}\fi{} M\'endez and Daniel Campos.
\newblock Characterization of stationary states in random walks with stochastic
  resetting.
\newblock {\em Phys. Rev. E}, 93:022106, Feb 2016.

\bibitem{Gupta_2019}
Deepak Gupta.
\newblock Stochastic resetting in underdamped brownian motion.
\newblock {\em Journal of Statistical Mechanics: Theory and Experiment},
  2019(3):033212, mar 2019.

\bibitem{Deepak-SL}
Deepak Gupta and Daniel~M. Busiello.
\newblock Tighter thermodynamic bound on the speed limit in systems with
  unidirectional transitions.
\newblock {\em Phys. Rev. E}, 102:062121, Dec 2020.

\bibitem{21Pal}
Arnab Pal, Shlomi Reuveni, and Saar Rahav.
\newblock Thermodynamic uncertainty relation for systems with unidirectional
  transitions.
\newblock {\em Phys. Rev. Research}, 3:013273, Mar 2021.

\bibitem{21Busiello}
Daniel~Maria Busiello, Deepak Gupta, and Amos Maritan.
\newblock Inducing and optimizing markovian mpemba effect with stochastic
  reset.
\newblock {\em New Journal of Physics}, 23(10):103012, oct 2021.

\bibitem{Ramoso}
A.~M. Ramoso, J.~A. Magalang, D.~S{\'{a}}nchez-Taltavull, J.~P. Esguerra, and
  {\'{E}}.~Rold{\'{a}}n.
\newblock Stochastic resetting antiviral therapies prevent drug resistance
  development.
\newblock {\em Europhysics Letters}, 132(5):50003, dec 2020.

\bibitem{Genthon_2022}
Arthur Genthon, Reinaldo Garc{\'{\i}}a-Garc{\'{\i}}a, and David Lacoste.
\newblock Branching processes with resetting as a model for cell division.
\newblock {\em Journal of Physics A: Mathematical and Theoretical},
  55(7):074001, jan 2022.

\bibitem{Santra_2022}
Ion Santra.
\newblock Effect of tax dynamics on linearly growing processes under stochastic
  resetting: A possible economic model.
\newblock {\em Europhysics Letters}, 137(5):52001, mar 2022.

\bibitem{11Evans_b}
Martin~R Evans and Satya~N Majumdar.
\newblock Diffusion with optimal resetting.
\newblock {\em Journal of Physics A: Mathematical and Theoretical},
  44(43):435001, oct 2011.

\bibitem{20Pal}
Arnab Pal, \L{}ukasz Ku\ifmmode~\acute{s}\else \'{s}\fi{}mierz, and Shlomi
  Reuveni.
\newblock Search with home returns provides advantage under high uncertainty.
\newblock {\em Phys. Rev. Research}, 2:043174, Nov 2020.

\bibitem{Pal_FPT}
Arnab Pal and Shlomi Reuveni.
\newblock First passage under restart.
\newblock {\em Phys. Rev. Lett.}, 118:030603, Jan 2017.

\bibitem{20Plata}
Carlos~A. Plata, Deepak Gupta, and Sandro Azaele.
\newblock Asymmetric stochastic resetting: Modeling catastrophic events.
\newblock {\em Phys. Rev. E}, 102:052116, Nov 2020.

\bibitem{14Reuveni}
Shlomi Reuveni, Michael Urbakh, and Joseph Klafter.
\newblock Role of substrate unbinding in michaelis\&\#x2013;menten enzymatic
  reactions.
\newblock {\em Proceedings of the National Academy of Sciences},
  111(12):4391--4396, 2014.

\bibitem{20Evans}
Martin~R Evans, Satya~N Majumdar, and Gr{\'{e}}gory Schehr.
\newblock Stochastic resetting and applications.
\newblock {\em Journal of Physics A: Mathematical and Theoretical},
  53(19):193001, apr 2020.

\bibitem{17Roldan}
\'Edgar Rold\'an and Shamik Gupta.
\newblock Path-integral formalism for stochastic resetting: Exactly solved
  examples and shortcuts to confinement.
\newblock {\em Phys. Rev. E}, 96:022130, Aug 2017.

\bibitem{18Chechkin}
A.~Chechkin and I.~M. Sokolov.
\newblock Random search with resetting: A unified renewal approach.
\newblock {\em Phys. Rev. Lett.}, 121:050601, Aug 2018.

\bibitem{18Evans}
Martin~R Evans and Satya~N Majumdar.
\newblock Effects of refractory period on stochastic resetting.
\newblock {\em Journal of Physics A: Mathematical and Theoretical},
  52(1):01LT01, nov 2018.

\bibitem{19MPuigdellosas}
Axel Mas\'o-Puigdellosas, Daniel Campos, and Vicen\ifmmode
  \mbox{\c{c}}\else~\c{c}\fi{} M\'endez.
\newblock Transport properties of random walks under stochastic
  noninstantaneous resetting.
\newblock {\em Phys. Rev. E}, 100:042104, Oct 2019.

\bibitem{20Mercado}
Gabriel Mercado-V{\'{a}}squez, Denis Boyer, Satya~N Majumdar, and Gr{\'{e}}gory
  Schehr.
\newblock Intermittent resetting potentials.
\newblock {\em Journal of Statistical Mechanics: Theory and Experiment},
  2020(11):113203, nov 2020.

\bibitem{21Santra}
Ion Santra, Santanu Das, and Sujit~Kumar Nath.
\newblock Brownian motion under intermittent harmonic potentials.
\newblock {\em Journal of Physics A: Mathematical and Theoretical},
  54(33):334001, jul 2021.

\bibitem{19Pal}
Arnab Pal, {\L}ukasz Ku{\'{s}}mierz, and Shlomi Reuveni.
\newblock Invariants of motion with stochastic resetting and space-time coupled
  returns.
\newblock {\em New Journal of Physics}, 21(11):113024, nov 2019.

\bibitem{20Brodova}
Anna~S. Bodrova and Igor~M. Sokolov.
\newblock Resetting processes with noninstantaneous return.
\newblock {\em Phys. Rev. E}, 101:052130, May 2020.

\bibitem{20Gupta}
Deepak Gupta, Carlos~A Plata, Anupam Kundu, and Arnab Pal.
\newblock Stochastic resetting with stochastic returns using external trap.
\newblock {\em Journal of Physics A: Mathematical and Theoretical},
  54(2):025003, dec 2020.

\bibitem{20Bressloff}
Paul~C. Bressloff.
\newblock Queueing theory of search processes with stochastic resetting.
\newblock {\em Phys. Rev. E}, 102:032109, Sep 2020.

\bibitem{21Zhou}
Tian Zhou, Pengbo Xu, and Weihua Deng.
\newblock Gaussian process and l\'evy walk under stochastic noninstantaneous
  resetting and stochastic rest.
\newblock {\em Phys. Rev. E}, 104:054124, Nov 2021.

\bibitem{20Tal}
Ofir Tal-Friedman, Arnab Pal, Amandeep Sekhon, Shlomi Reuveni, and Yael
  Roichman.
\newblock Experimental realization of diffusion with stochastic resetting.
\newblock {\em The Journal of Physical Chemistry Letters}, 11(17):7350--7355,
  2020.
\newblock PMID: 32787296.

\bibitem{21Faisant}
F~Faisant, B~Besga, A~Petrosyan, S~Ciliberto, and Satya~N Majumdar.
\newblock Optimal mean first-passage time of a brownian searcher with resetting
  in one and two dimensions: experiments, theory and numerical tests.
\newblock {\em Journal of Statistical Mechanics: Theory and Experiment},
  2021(11):113203, nov 2021.

\bibitem{10sekimoto}
Ken Sekimoto.
\newblock {\em Stochastic {Energetics}}.
\newblock Springer, March 2010.

\bibitem{15Meylahn}
Janusz~M. Meylahn, Sanjib Sabhapandit, and Hugo Touchette.
\newblock Large deviations for markov processes with resetting.
\newblock {\em Phys. Rev. E}, 92:062148, Dec 2015.

\bibitem{16Fuchs}
Jaco Fuchs, Sebastian Goldt, and Udo Seifert.
\newblock Stochastic thermodynamics of resetting.
\newblock {\em {EPL} (Europhysics Letters)}, 113(6):60009, mar 2016.

\bibitem{Deepak_PRL}
Deepak Gupta, Carlos~A. Plata, and Arnab Pal.
\newblock Work fluctuations and jarzynski equality in stochastic resetting.
\newblock {\em Phys. Rev. Lett.}, 124:110608, Mar 2020.

\bibitem{Deepak_PRR}
D.~M. Busiello, D.~Gupta, and A.~Maritan.
\newblock Entropy production in systems with unidirectional transitions.
\newblock {\em Phys. Rev. Research}, 2:023011, Apr 2020.

\bibitem{17Pal}
Arnab Pal and Saar Rahav.
\newblock Integral fluctuation theorems for stochastic resetting systems.
\newblock {\em Phys. Rev. E}, 96:062135, Dec 2017.

\bibitem{13Evans}
Martin~R Evans, Satya~N Majumdar, and Kirone Mallick.
\newblock Optimal diffusive search: nonequilibrium resetting versus equilibrium
  dynamics.
\newblock {\em Journal of Physics A: Mathematical and Theoretical},
  46(18):185001, apr 2013.

\bibitem{Gupta_2021_TD}
Deepak Gupta, Arnab Pal, and Anupam Kundu.
\newblock Resetting with stochastic return through linear confining potential.
\newblock {\em Journal of Statistical Mechanics: Theory and Experiment},
  2021(4):043202, apr 2021.

\bibitem{01Redner_book}
Sidney Redner.
\newblock {\em A Guide to First-Passage Processes}.
\newblock Cambridge University Press, 2001.

\end{thebibliography}

\end{document}